\documentclass{iau307}
\usepackage{graphicx}
\usepackage{natbib}
\usepackage{url}
\usepackage{dtklogos}
\bibpunct{(}{)}{;}{a}{}{,}

\title[Semi-Convection in $\beta$ Cep Stars] 
{Asteroseismic Diagnostics for Semi-Convection in B Stars in the Era of K2}

\author[Ehsan Moravveji]   
{Ehsan Moravveji$^1$
\thanks{Postdoctoral Fellow of the Belgian Science Policy Office (BELSPO), Belgium}}

\affiliation{$^1$Instituut voor Sterrenkunde, KU Leuven, Celestijnenlaan 200D, B-3001 Leuven, Belgium \\ 
email: {\tt Ehsan.Moravveji@ster.kuleuven.be} \\[\affilskip] }

\pubyear{2014}
\volume{307} 
\pagerange{}
\jname{New windows on massive stars: asteroseismology, interferometry, and spectropolarimetry}
\editors{G. Meynet, C. Georgy, J.H. Groh \& Ph. Stee, eds.}

\begin{document}

\maketitle

\begin{abstract}
  Semi-convection is a slow mixing process in chemically-inhomogeneous radiative
  interiors of stars.  In massive OB stars, it is important during the main
  sequence.  However, the efficiency of this mixing mechanism is not properly
  gauged yet.  Here, we argue that asteroseismology of $\beta$ Cep pulsators is
  capable of distinguishing between models of varying semi-convection
  efficiencies.  We address this in the light of upcoming high-precision space
  photometry to be obtained with the \textit{Kepler} two-wheel mission for
  massive stars along the ecliptic.

\end{abstract}

\firstsection 

\section{Introduction}\label{s-intro}
Non-radial pulsation is a common phenomenon among B dwarfs.  Early-type B stars
- widely known as $\beta$ Cep stars - are pulsationally unstable against
low-order, low-degree radial and non-radial pressure (p-) and gravity (g-)
modes.  Their mass ranges from $\sim$8 to 20 M$_\odot$ \citep[see][for
details]{aerts-2010-book}.  Contrary to their fully mixed convective cores, the
mixing of species in their radiative interior occurs on a long - yet unconstrained
- time scale.  There are several mixing mechanisms that operate (simultaneously)
in radiative zones, among which rotational mixing and semi-convection.  In this
paper, we limit ourselves to slowly-rotating B stars.  

The pulsation frequencies of stars are highly sensitive to their internal
structure, and can be used as a proxy to test different input physics.
\cite{miglio-2008-01} already showed the effect of extra mixing induced by, e.g.,
rotation, atomic diffusion, and convective overshooting on the period spacing of
g-modes in heat-driven pulsators; their conclusions can be extended to include
semi-convection as an extra mixing mechanism.

In the near future, the \textit{Kepler} two-wheel mission, \citep[hereafter
K2,][]{howell-2014-01} will provide high-precision space photometry of a handful
of late-O and early B-type pulsators in the ecliptic plane.  We emphasise that
K2 will conduct pioneering observations, since such space photometry is scarce
for massive stars, particularly for objects more massive than $8\,$M$_\odot$.

In this paper, we put forward asteroseismic diagnostics to probe semi-convective
mixing in massive main-sequence stars.  Following on \cite{miglio-2008-01}, we
address the possibility of constraining the efficiency of semi-convection in
massive stars in light of the upcoming high-precision data to be assembled by
the K2 mission and already present in the CoRoT archive.

\section{Semi-Convective Mixing}\label{s-Sch-Led}

Semi-convection is a slow mixing process believed to operate in the
chemically inhomogeneous parts of radiative zones, where the g-modes are
oscillatory \citep{schwarzschild-1958-01, kato-1966-01, langer-1983-01,
  noels-2010-01}.  In other words, semi-convection acts in those layers of the
star where the radiative temperature gradient $\nabla_{\rm rad}$ takes values in
between the
adiabatic  $\nabla_{\rm ad}$ and the Ledoux $\nabla_{\rm L}$ gradients,
i.e., $\nabla_{\rm ad}<\nabla_{\rm rad}<\nabla_{\rm L}$.  Here, $\nabla_{\rm
  L}=\nabla_{\rm ad}+\varphi/\delta \,\nabla_\mu$, with
$\varphi=(\partial\ln\rho/\partial\ln T)_{P,\mu}$,
$\delta=(\partial\ln\rho/\partial\ln\mu)_{P,T}$ and $\nabla_\mu=d\ln\mu/d\ln P$.
Semi-convection occurs due to 
the stabilizing effect of the composition gradient $\nabla_\mu$
against the onset of convection in the chemically inhomogeneous layers on top of
the receding convective core.  This mixing process is believed to be present
in stars with $M\gtrsim15$ M$_\odot$ \citep{langer-1985-01, langer-1991-01}.
The reason is the increasing effect of the radiation pressure with mass, and the
local increase of $\nabla_{\rm rad}$ with respect to $\nabla_{\rm ad}$ outside the
convective core.  Therefore, semi-convection is not expected in intermediate to
late B-type stars.  For this reason, $\beta$ Cep stars are optimal candidates to
investigate if such a mixing can leave observable footprints.  The
semi-convective mixing is typically described in a diffusion approximation
\citep{langer-1983-01}.  See Section 6.2 in
\citet{maeder-2009-book} for an overview of different mixing schemes.

\begin{figure}
\begin{minipage}{0.50\textwidth}
\includegraphics[width=\columnwidth]{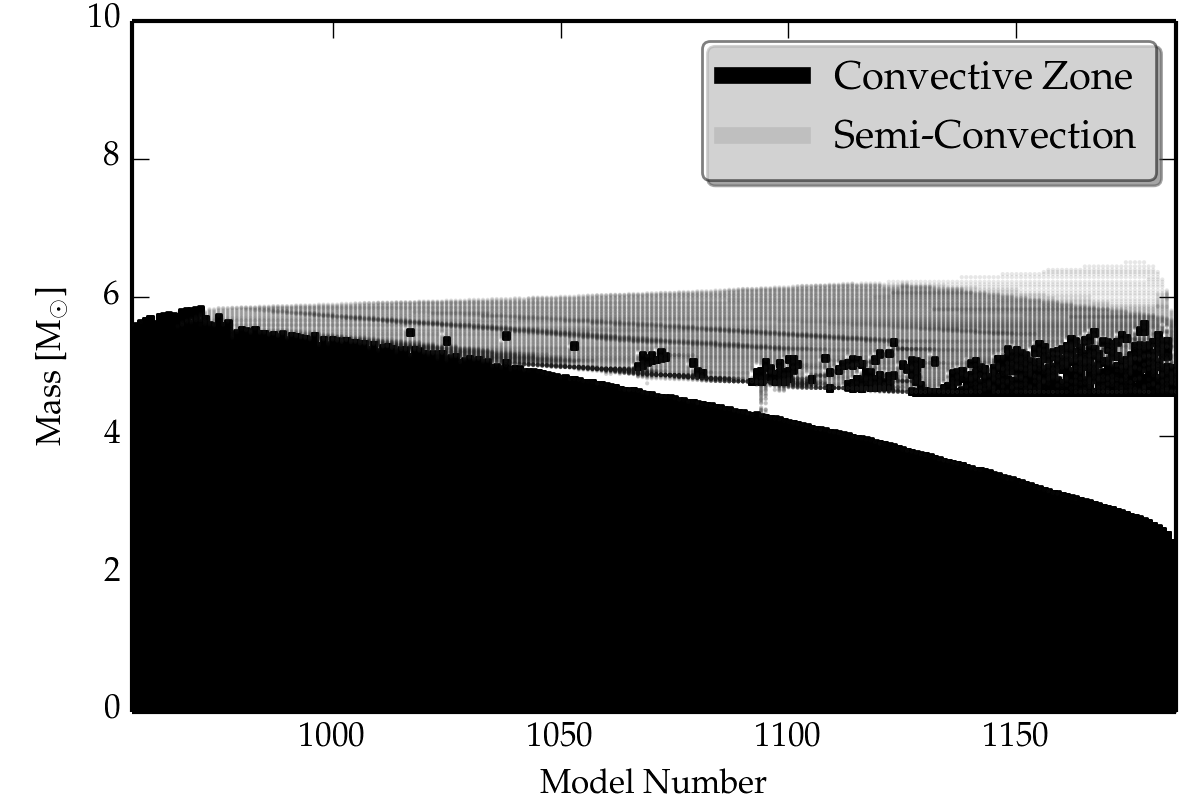}
\end{minipage}
\begin{minipage}{0.50\textwidth}
\includegraphics[width=\columnwidth]{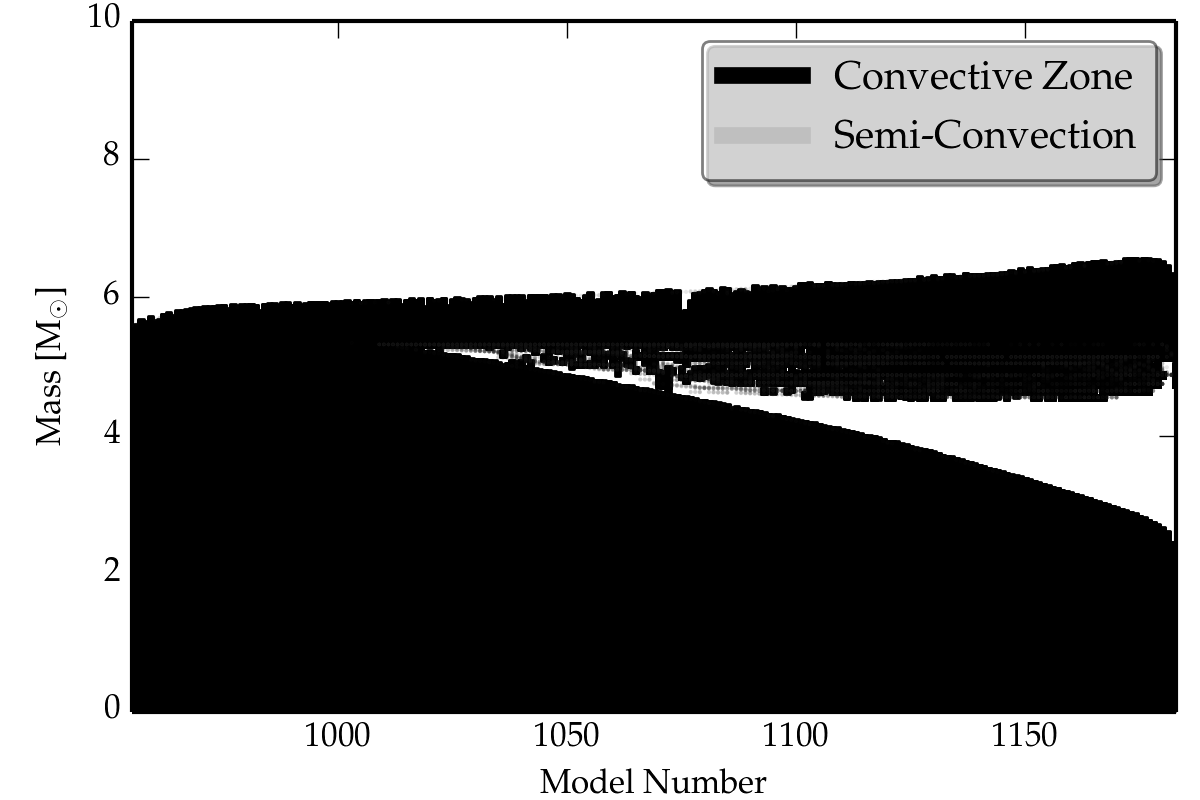}
\end{minipage}
\caption{Kippenhahn diagrams showing the evolution of 15 M$_\odot$ stellar
  models with $\alpha_{\rm sc}=10^{-1}$ (left) and $\alpha_{\rm sc}=1$ (right).
  The convective zones are shown in blue, and the semi-convective zones are
  shown in purple.}\label{f-Kipp}
\end{figure}

The stellar evolution code
MESA \citep{paxton-2011-01, paxton-2013-01} follows the prescription by \cite{langer-1985-01}, where the semi-convective 
diffusion coefficient $D_{\rm sc}$ is defined according to  
\citet{kato-1966-01} and \citet{langer-1983-01}:
\begin{equation}\label{e-sc}
D_{\rm sc}=\alpha_{\rm sc} \frac{\kappa_r}{6c_p\rho}\frac{\nabla-\nabla_{\rm
    ad}}{\nabla_{\rm L}-\nabla},
\end{equation}
where $\kappa_r=4acT^3/3\kappa\rho$ is the radiative conductivity, $c_p$ the
specific heat at constant pressure, and $\rho$ denotes the density.  The
efficiency of semi-convective mixing is controlled by the free parameter
$\alpha_{\rm sc}$, which determines the length scale and time scale of the vibrational 
mixing associated to semi-convective zones.  
The parameter $\alpha_{\rm sc}$ is not calibrated from
observations, \citet[e.g.,][]{langer-1985-01} favoured $\alpha_{\rm sc}=10^{-1}$,
while later on \cite{langer-1991-01} preferred $0.01\leq\alpha_{\rm
  sc}\leq0.04$.  Note that the value of $\alpha_{\rm sc}$ depends on the choice
of the opacity tables and on the numerical scheme to compute $D_{\rm sc}$ according
to Eq.\,(\ref{e-sc}).  We aim at constraining $\alpha_{\rm sc}$ using K2 data.

\section{Effect of Semi-Convection on $\mathbf{\beta}$ Cep models}\label{s-mesa}
We used the MESA code to calculate four evolutionary tracks for a 15 M$_\odot$
star with initial chemical composition $(X,Y,Z)=(0.710, 0.276, 0.014)$ based on
\cite{nieva-2012-01}.  We excluded mass loss in all models and considered four
values of $\alpha_{\rm sc}=10^{-6}$, $10^{-4}$, $10^{-2}$, and 1.  We used the
Ledoux criterion of convection, and made sure that the condition $\nabla_{\rm
  rad}=\nabla_{\rm ad}=\nabla_{\rm L}$ was satisfied from the
convective side of the core boundary \citep{gabriel-2014-01}; see also Noels
(these proceedings).  On each evolutionary track, we stored an equilibrium model
for a central hydrogen abundance of $X_c=0.10$.  We subsequently used the GYRE
pulsation code \citep{townsend-2013-01} to calculate radial $\ell=0$, dipole
$\ell=1$ and quadrupole $\ell=2$ mode frequencies in the adiabatic
approximation, for each model.  We restricted the comparison of the mode
behaviour to low-order modes, i.e., $-5\leq n_{\rm pg}\leq+3$, where $n_{\rm
  pg}=n_{\rm p}-n_{\rm g}$.

Figure \ref{f-Kipp} shows Kippenhahn diagrams for models with $\alpha_{\rm
  sc}=10^{-2}$ (left) and $\alpha_{\rm sc}=1$ (right).  For relatively limited
semi-convective mixing (i.e., $\alpha_{\rm sc}=10^{-2}$), an extended
semi-convective zone (grey points) develops and continues at roughly the same
mass coordinate.  On the other hand, for efficient semi-convective mixing
(i.e. $\alpha_{\rm sc}=1$) the former zones are identified as convective and an
extended intermediate convective zone (hereafter ICZ) develops.  In our models,
the ICZ encapsulates $\sim$ 1 M$_\odot$.  The formation and presence of an ICZ
largely impacts the later evolution of the star.  From an
asteroseismic point of view, models harbouring an ICZ have smaller radiative
zones (measured from the surface), hence a smaller cavity for g-mode
propagation.  As a result, it is expected that the presence of an ICZ, which in
turn results from different semi-convective efficiencies, affects the adiabatic
frequencies of low-order low-degree p- and g-modes.

\begin{figure}[t!]
\centering
\includegraphics[width=\columnwidth]{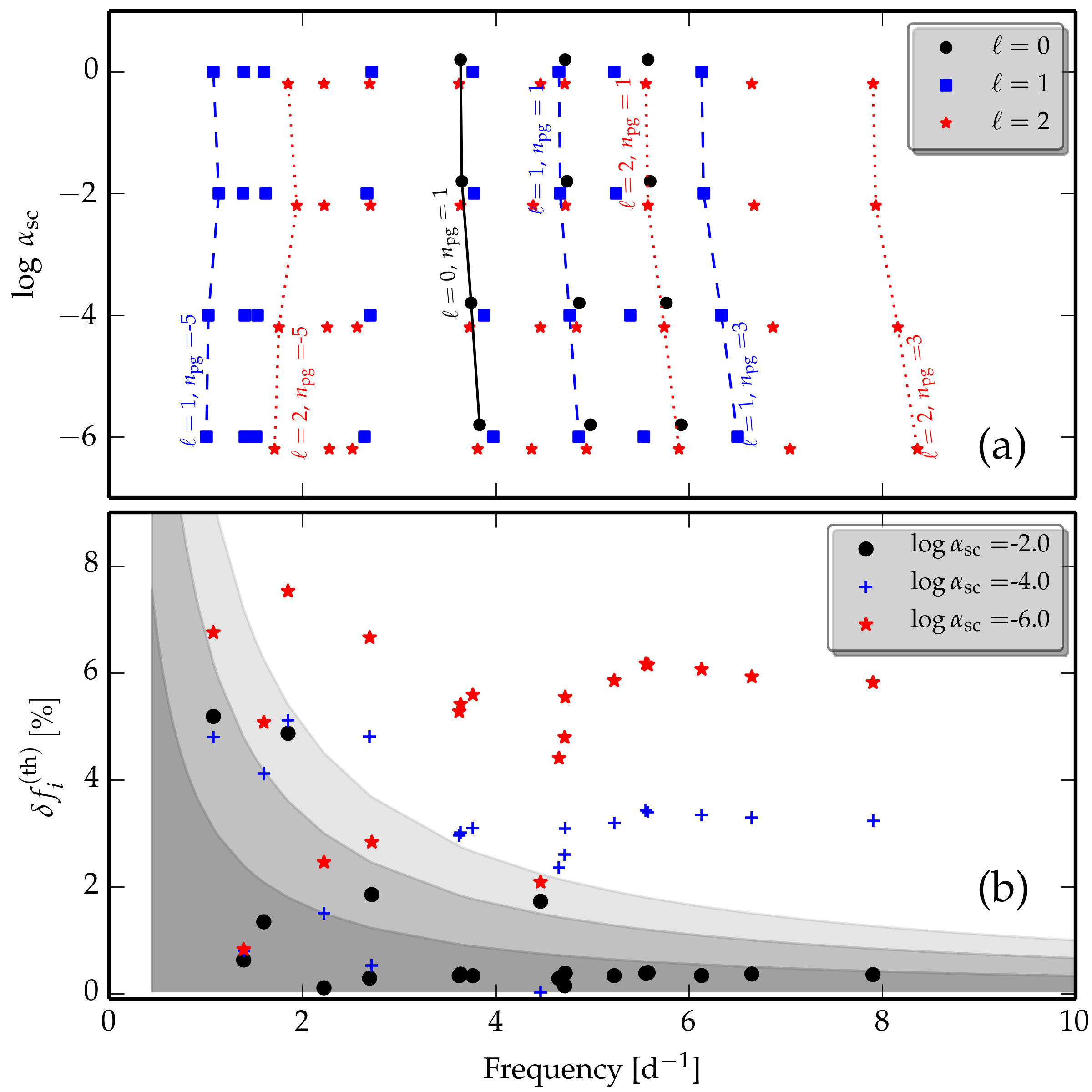}
\caption{
  Adiabatic low radial order $-5\leq n_{\rm pg}\leq+3$ frequencies of 15
  M$_\odot$ models with Z=Z$_\odot$ and $X_c=0.10$.  Both panels share the same
  frequency range on the abscissa.  (a) The ordinate is the logarithm of
  $\alpha_{\rm sc}$ (Eq. \ref{e-sc}).  Circles are radial $\ell=0$ modes,
  squares are non-radial dipole $\ell=1$ p- and g-modes, and stars are
  non-radial quadrupole $\ell=2$ p- and g-modes.  (b) Comparison of the relative
  theoretical frequency change $\delta f^{\rm(th)}$ with the relative observed
  frequency precision $\delta f^{\rm(obs)}$ as a function of frequency; see
  Eqs. (\ref{e-delta-f-th}) and (\ref{e-delta-f-obs}).  Dark grey, grey and
  light grey show 1$\sigma$, 2$\sigma$ and 3$\sigma$ precision levels,
  respectively.  Circles, plus marks, and stars correspond to models with
  $\alpha_{\rm sc}=10^{-2}, \,10^{-4}$, and $10^{-6}$, respectively.  The
  reference frequencies are taken from the $\alpha_{\rm sc}=1$ model.
  Consult the color version of this figure in electronic format.}
  \label{f-freq-asc}
\end{figure}

A word of caution about matching detected frequencies of unidentified modes from
a grid of asteroseismic models is worthwhile to be made here.  Figure
\ref{f-freq-asc}a shows that the frequency of the radial fundamental mode
$\ell=0, n_{\rm pg}=1$ is quite close to that of an $\ell=2, n_{\rm pg}=-2$
non-radial g-mode.  Without robust mode identification --- which by itself is
intricate --- the interpretation of detected pulsation frequencies can be quite
misleading.  Therefore, the photometric light curves to be assembled by K2 must
be complemented with ground-based multi-colour photometry and/or high-resolution
spectroscopy to identify at least one of the detected modes.  Non-adiabatic
computations might help to identify modes, although there are still severe
disagreements between the observed modes in OB stars and those predicted to be
excited in the sense that we detect many more modes than foreseen by theory. It
is therefore safer not to rely on excitation computations when identifying
detected frequencies.

Figure \ref{f-freq-asc}a shows the frequencies of $\ell=0$, 1 and 2 p- and
g-modes for different values of $\alpha_{\rm sc}$.  For better visibility, few
modes with identical $\ell$ and $n_{\rm pg}$ are connected by lines.  Clearly,
the change in $\alpha_{\rm sc}$ shifts most of the frequencies to slightly
lower/higher values.  The frequency difference of the modes with fixed $\ell$
and $n_{\rm pg}$ increases with decreasing $\alpha_{\rm sc}$.  An important question
is whether we are able to capture such subtle differences observationally from 
the K2 space photometry.

To answer that, we take the model with $\alpha_{\rm sc}=1$ as the
\textit{reference} model, hence its frequencies are $f_i^{\rm (ref)}$.  We
compare the relative frequency change $\delta f_i^{\rm (th)}$ of each
theoretical frequency $f_i$ with respect to the reference frequency as
\begin{equation}\label{e-delta-f-th}
\delta f_i^{\rm (th)} = \frac{|f_i^{\rm (ref)} - f_i|}{f_i^{\rm (ref)} }.
\end{equation}
Here, $f_i^{\rm (ref)}$ and $f_i$ are both calculated using GYRE.  From an
observational point of view, the frequency precision $\Delta f$ depends on the
total observation time base $\Delta T$.  A conservative estimate by
\cite{loumos-1978-01} is $\Delta f \approx 2.5/\Delta T$.  For K2, the
planned $\Delta T$ is approximately 75 days.  We also define the relative
observed frequency precision $\delta f_i^{\rm (obs)}$
\begin{equation}\label{e-delta-f-obs}
\delta f_i^{\rm (obs)} = \frac{\Delta f}{f_i^{\rm (ref)}},
\end{equation}
where $\delta f_i^{\rm (obs)}$ stands for a 1$\sigma$ uncertainty level, and
measures the observational frequency precision required to capture the effect of
a specific feature --- in our case the presence/absence of the ICZ.  If by varying
a stellar structure free parameter --- here $\alpha_{\rm sc}$ --- the relative
theoretical frequency change is significantly larger than the estimated
observational relative frequency precision, i.e. $\delta f_i^{\rm (th)} \gtrsim 3 \,\delta
f_i^{\rm (obs)}$, then asteroseismology can 
constrain the value of that parameter.
Note the arbitrary choice of $3\sigma$ here.

To visualise the probing power of the K2 data, Figure \ref{f-freq-asc}b shows
the relative frequency change $\delta f_i^{\rm (th)}$ with respect to the
reference model ($\alpha_{\rm sc}=1$) for models with $\log\alpha_{\rm sc}=-2$
(circles), $-$4 (plus marks) and $-$6 (stars), respectively.  In the background of
the same plot, we show $\delta f_i^{\rm (obs)}$ (dark grey), $2\times\delta
f_i^{\rm (obs)}$ (grey) and $3\times\delta f_i^{\rm (obs)}$ (light grey),
respectively.  The distribution of the circles is roughly inside the 1$\sigma$ zone,
which makes the distinction between models with $\alpha_{\rm sc}=1$
and $\alpha_{\rm sc}=10^{-2}$ quite challenging.  However, plus marks that compare
$\log\alpha_{\rm sc}=10^{-4}$ models with those of the reference model, or stars
that compare $\log\alpha_{\rm sc}=10^{-6}$ models with the reference model are 
more than 3$\sigma$ away from $\delta f^{\rm (obs)}$.  
For the latter, K2 holds the potential to constrain the efficiency of
semi-convection as an extra mixing mechanism, provided that we can find good
seismic models of the stars according to the scheme outlined in Aerts (these
proceedings) to which we then add the concept of semi-convective mixing.

\section{Conclusions}\label{s-conclude}
Unfortunately, the \textit{Kepler\/} mission observed no O-type and early-B
stars so far.  Thus, we are yet unable to estimate the richness of the frequency
spectrum of the K2 light curves for massive dwarfs.  The public release of the
K2 field 0 light curves is scheduled for September 2014; since there are several
OB pulsators on K2 silicon, we will soon be able to gauge the quality of K2 data 
for massive star asteroseismology.  

In our comparisons, we assumed that the K2 data do not suffer from instrumental
effects, hence that $\delta f^{\rm (obs)}$ depends only on the time base of the
K2 campaigns.  This is of course an idealised situation.  The telescope jitter
and drift can easily increase $\delta f^{\rm (obs)}$.  Yet, the seismic
diagnostic potential stays valid since the $\delta f^{\rm (th)}$ can exceed the
3$\sigma$ level for several modes, and hence the effect of semi-convective
mixing can hopefully be detected and studied using K2 light curves.

Based on Figure \ref{f-freq-asc}, there is no real preference between modes of
different order $n_{\rm pg}$ and degree $\ell$ in providing asteroseismic diagnostics for semi-convection; in
other words, all low-order p- and g-modes possess the same potential to provide
a constraint on $\alpha_{\rm sc}$.  The semi-convection free parameter was varied
in a broad range, from $10^{-6}$ to $1$, in our exercise. It is of course easier
to discriminate between models with large differences in $\alpha_{\rm
  sc}$.  According to Figure \ref{f-freq-asc}b, the distinction between models with
$\alpha_{\rm sc}=10^{-6}$ and 1 is more within reach than distinguishing
between models with $\alpha_{\rm sc}=10^{-2}$ and 1.  

The main message of our work is to emphasis the importance of semi-convection
along with the shrinking convective cores in massive OB stars, when the observed
frequencies of $\beta$ Cep stars are compared with theoretical frequencies.
Consequently, semi-convection --- as one of the extra mixing mechanisms in
inhomogeneous layers of the stellar radiative interior --- could be employed as
an extra dimension when modelling stars based on grid calculations coupled to
asteroseismic forward modelling \citep{briquet-2007-01}.

\bibliographystyle{iau307}
\bibliography{/Users/ehsan/my/papers/my-bib.bib}

\begin{thebibliography}{}

\bibitem[\protect\astroncite{{Aerts} et~al.}{2010}]{aerts-2010-book}
{Aerts}, C., {Christensen-Dalsgaard}, J., \& {Kurtz}, D.~W. 2010,
\newblock {\em {Asteroseismology, Astronomy and Astrophsyics Library, Springer
  Berlin Heidelberg}}

\bibitem[\protect\astroncite{{Briquet} et~al.}{2007}]{briquet-2007-01}
{Briquet}, M., {Morel}, T., {Thoul}, A., {et~al.} 2007,
\newblock {\em \mnras} 381, 1482

\bibitem[\protect\astroncite{{Gabriel} et~al.}{2014}]{gabriel-2014-01}
{Gabriel}, M., {Noels}, A., {Montalban}, J., \& {Miglio}, A. 2014,
\newblock {\em ArXiv e-prints}

\bibitem[\protect\astroncite{{Howell} et~al.}{2014}]{howell-2014-01}
{Howell}, S.~B., {Sobeck}, C., {Haas}, M., {et~al.} 2014,
\newblock {\em \pasp} 126, 398

\bibitem[\protect\astroncite{{Kato}}{1966}]{kato-1966-01}
{Kato}, S. 1966,
\newblock {\em \pasj} 18, 374

\bibitem[\protect\astroncite{{Langer}}{1991}]{langer-1991-01}
{Langer}, N. 1991,
\newblock {\em \aap} 252, 669

\bibitem[\protect\astroncite{{Langer} et~al.}{1985}]{langer-1985-01}
{Langer}, N., {El Eid}, M.~F., \& {Fricke}, K.~J. 1985,
\newblock {\em \aap} 145, 179

\bibitem[\protect\astroncite{{Langer} et~al.}{1983}]{langer-1983-01}
{Langer}, N., {Fricke}, K.~J., \& {Sugimoto}, D. 1983,
\newblock {\em \aap} 126, 207

\bibitem[\protect\astroncite{{Loumos} \& {Deeming}}{1978}]{loumos-1978-01}
{Loumos}, G.~L. \& {Deeming}, T.~J. 1978,
\newblock {\em \apss} 56, 285

\bibitem[\protect\astroncite{{Maeder}}{2009}]{maeder-2009-book}
{Maeder}, A. 2009,
\newblock {\em {Physics, Formation and Evolution of Rotating Stars}}

\bibitem[\protect\astroncite{{Miglio} et~al.}{2008}]{miglio-2008-01}
{Miglio}, A., {Montalb{\'a}n}, J., {Noels}, A., \& {Eggenberger}, P. 2008,
\newblock {\em \mnras} 386, 1487

\bibitem[\protect\astroncite{{Mowlavi} \& {Forestini}}{1994}]{mowlavi-1994-01}
{Mowlavi}, N. \& {Forestini}, M. 1994,
\newblock {\em \aap} 282, 843

\bibitem[\protect\astroncite{{Nieva} \& {Przybilla}}{2012}]{nieva-2012-01}
{Nieva}, M.-F. \& {Przybilla}, N. 2012,
\newblock {\em \aap} 539, A143

\bibitem[\protect\astroncite{{Noels} et~al.}{2010}]{noels-2010-01}
{Noels}, A., {Montalban}, J., {Miglio}, A., {Godart}, M., \& {Ventura}, P.
  2010,
\newblock {\em \apss} 328, 227

\bibitem[\protect\astroncite{{P{\'a}pics} et~al.}{2014}]{papics-2014-01}
{P{\'a}pics}, P.~I., {Moravveji}, E., {Aerts}, C., {et~al.} 2014,
\newblock {\em \aap} , in press (arXiv1407.2986)

\bibitem[\protect\astroncite{{Paxton} et~al.}{2011}]{paxton-2011-01}
{Paxton}, B., {Bildsten}, L., {Dotter}, A., {et~al.} 2011,
\newblock {\em \apjs} 192, 3

\bibitem[\protect\astroncite{{Paxton} et~al.}{2013}]{paxton-2013-01}
{Paxton}, B., {Cantiello}, M., {Arras}, P., {et~al.} 2013,
\newblock {\em \apjs} 208, 4

\bibitem[\protect\astroncite{{Schwarzschild} \&
  {H{\"a}rm}}{1958}]{schwarzschild-1958-01}
{Schwarzschild}, M. \& {H{\"a}rm}, R. 1958,
\newblock {\em \apj} 128, 348

\bibitem[\protect\astroncite{{Townsend} \& {Teitler}}{2013}]{townsend-2013-01}
{Townsend}, R.~H.~D. \& {Teitler}, S.~A. 2013,
\newblock {\em \mnras} 435, 3406

\end{thebibliography}

\end{document}